\documentclass[prb,twocolumn,showpacs,showkeys,superscriptaddress]{revtex4}%
\usepackage{times}
\usepackage{graphicx}
\usepackage{amsmath}
\usepackage{amsfonts}
\usepackage{amsthm}
\usepackage{amssymb}
\usepackage{mathrsfs}

\begin{document}

\title{Open quantum system description of singlet-triplet qubits in quantum dots}

\author{L. K. Castelano}
\email{lkcastelano@ufscar.br}
\affiliation{Departamento de F\'{\i}sica, Universidade Federal de S\~ao Carlos, 13565-905, S\~ao Carlos, SP, Brazil}
\author{F. F. Fanchini}
\affiliation{Faculdade de Ci\^encias, UNESP - Universidade Estadual Paulista, Bauru, SP, 17033-360, Brazil}
\affiliation{The Abdus Salam International Centre for Theoretical Physics, Strada Costiera 11, Miramare-Trieste, Italy}
\author{K. Berrada}
\affiliation{The Abdus Salam International Centre for Theoretical Physics, Strada Costiera 11, Miramare-Trieste, Italy}
\affiliation{Physics Department, College of Science, Al Imam Mohammad Ibn Saud Islamic University (IMISU), Riyadh, Saudi Arabia.}

\begin{abstract}
We develop a theoretical model to describe the dissipative dynamics of singlet-triplet qubits in GaAs quantum dots. Using the concurrence experimentally obtained [M. D. Shulman \textit{et al.}, Science \textbf{336}, 202 (2012)] as a guide, we found that each logical qubit fluctuates under the action of a random telegraph noise (RTN) that simulates the 1/f$^\alpha$ noise. We also study the dynamics of concurrence as a function of the amplitude of the RTN, the correlation time of the RTN, the preparation time of states, and the two-qubit coupling. Furthermore, we show that the two-qubit coupling together with the preparation time strongly affect the entanglement dissipative dynamics and both physical quantities can be employed to enhance the entanglement between singlet-triplet qubits.

\end{abstract}

\date{\today}
 \maketitle

\section{Introduction}
The development of quantum information processing has enabled the discovery of new techniques and platforms which are paving the way to accomplish quantum technologies in the near future.~ \cite{newt}
Among these platforms, spin qubits in quantum dots (QDs)\cite{lossdi} is certainly one of the most striking systems because of their potential scalability and miniaturization.~\cite{scale1,scale2,scale3} Furthermore, electrical readout and control of spins qubits in QDs have been achieved in several different approaches,~\cite{loss}
where spin blockade and charge sensors enable the observation of single/two-spin dynamics.~\cite{qd2} In double quantum dots (DQDs), a logical qubit can be encoded by means of singlet-triplet ($S$-$T_0$) states of two electron spins~\cite{qd1,levy,qdn1} and
the inter-qubit interaction can be implemented through a capacitive coupling.~\cite{weperen} By controling and coupling $S$-$T_0$ qubits, the entanglement between two $S$-$T_0$ qubits has been experimentally demonstrated.~\cite{sci} Together with the success of such a demonstration, the ubiquitous noise has been probed in the experimental data of ref.~[\onlinecite{sci}]. For a single $S$-$T_0$ qubit, the noise has been characterized and shown to be consistent with the power-law 1/f$^\alpha$ noise.~\cite{prl_shulman} Also, it has been experimentally verified that the expoent $\alpha$ has a temperature dependence; for instance, $\alpha\approx 0.7$ for T=50~mK and $\alpha\approx 0$ for T= 100~mK.~\cite{prl_shulman} Such a temperature dependence can be ascribed to phonon-induced decoherence mechanism.~\cite{phonon} 
The 1/f$^\alpha$ noise can be present in a variety of systems~\cite{rmp} and particullarly in other QDs systems.~\cite{chargenoise1,chargenoise2} To model the 1/f$^\alpha$ noise, random telegraph noise (RTN) has been employed in different theoretical works.~\cite{artigoRTN,guido,twortn,pra}

In this work, we employ the RTN to describe the decoherence caused by the interaction between two $S$-$T_0$ qubits and their environment in the low temperature limit, \textit{i.e.} $\alpha\neq 0$. By using such a model, we are able to quantitatively reproduce experimental results obtained for $S$-$T_0$ qubits in two GaAs coupled DQDs.~\cite{sci} Through our description of this open quantum system, we exploit the role of the amplitude of the RTN, the correlation time of the RTN, the preparation time of states, and the two-qubit coupling in the entanglement dissipative dynamics.

The present paper is organized as follows. In the next section, we present the model that describes the dynamics of the open quantum system. In Sec. III, we introduce the concept of entanglement, measured by concurrence, together with results of our theoretical model. We also include in Sec. III a detailed study on the most important physical parameters that rules the entanglement dynamics. Finally, Sec. IV contains a summary of our results.

\section{Theoretical Model}
The main focus of this work is related to the study of the dissipative dynamics of two $S$-$T_0$ qubits, where the information is stored in the spin states of two electrons. Such states can be experimentally achieved by confining two electrons in each DQD system.~\cite{sci} The logical qubit composed by the two-level system ($|S\rangle\equiv|\uparrow\rangle$,$|T_0\rangle\equiv|\downarrow\rangle$) can be isolated by applying an external magnetic field in the plane of the device in such a way that the Zeeman splitting makes the parallel spin states $|T_+\rangle$ and $|T_-\rangle $ energetically inaccessible.

To extend such a two-level system to a two-qubit system, it is necessary to couple two $S$-$T_0$ qubits, where the tunnelling between them is suppressed and their coupling is electrostatic (for more details, see ref.~[\onlinecite{sci}]). Thus, the effective Hamiltonian for the two-qubit system can be written as follows:~\cite{sci}     
\begin{eqnarray}
\hat{H}_{\text{2-qubit}}&=&{\dfrac{1}{2}}\left(J_1\sigma_z^{(1)}\otimes \mathbf{I}+J_2\;\mathbf{I}\otimes \sigma_z^{(2)}\right)+\nonumber\\ &&{\dfrac{J_{12}}{4}}\left(\sigma_z^{(1)}\otimes
\sigma_z^{(2)}-\sigma_z^{(1)}\otimes \mathbf{I}-\mathbf{I}\otimes\sigma_z^{(2)}\right)+ \nonumber\\ &&{\dfrac{1}{2}}\left(\Delta B_{z,1}\sigma_x^{(1)}\otimes\mathbf{I} +\Delta B_{z,2}\;\mathbf{I}\otimes \sigma_x^{(2)}\right),\label{ham}
\end{eqnarray}
where $\sigma_{x,y,z}$ are the Pauli spin matrices, $\mathbf{I}$ is the identity and the index 1 (2) is related to the first (second) qubit (hereafter, we use units of $\hbar =1$). This Hamiltonian is able to implement universal quantum control, which is given by two physically distinct local operations, $x$ and $z$, and by the interaction between the qubits given by $\sigma_z^{(1)}\otimes\sigma_z^{(2)}$. The exchange splitting, $J_{i}$, between $|S^i\rangle$ and $|T^i_0\rangle$ applies rotations in the qubit $i$=1,2 around the $z$ axis, while rotations around the $x$ axis are driven by a magnetic field gradient $\Delta B_z$.  Moreover, $\Delta B_z$ is responsible for the preparation of each qubit in a superposition between $|S\rangle$ and $|T_0\rangle$. The two-qubit coupling, $J_{12}$, depends on the energy between levels of the left and the right DQD and it can be switched on and off during the quantum dynamics.~\cite{sci} Due to the Pauli exclusion principle, $|S\rangle$ and $|T_0\rangle$ states have different charge configurations and because both qubits are electrostatically coupled, the state of the first qubit is conditioned to the state of the second qubit. In other words, when simultaneously evolving, they experience a dipole-dipole coupling that generates an entangled state. Following the experimental steps,~\cite{sci} each qubit is initialized in the $|S\rangle$ state, then rotated by $\pi/2$ around the $x$ axis when $J_i=J_{12}=0$, $\Delta B_{z,i}/2\pi\approx 30 \rm {MHz}$, for i=1,2. After this stage, a large exchange splitting is switched on corresponding to $J_1/2\pi\approx 280 {\rm {MHz}}$, and $J_2/2\pi\approx 320 {\rm {MHz}}$. Experimentally, it was found that the two-qubit coupling is given by $J_{12}= J_1J_2$.~\cite{sci}

To include the dissipative dynamics, we consider a phenomenological approach, where
both qubits are subjected to local RTN fluctuations on  exchange splitting terms $J_1$ and $J_2$. Thus, the RTN-Hamiltonian can be written as
\begin{eqnarray}
\hat{H}_{\text{RTN}}&=& J^{\text{RTN}}_1(t,\tau_c)\sigma_z^{(1)}\otimes \mathbf{I}+J^{\text{RTN}}_2(t,\tau_c)\mathbf{I}\otimes \sigma_z^{(2)}.\label{hamRTN}
\end{eqnarray}
For such a kind of noise, $J^{\text{RTN}}_k(t,\tau_c)$  jumps between two values $-J_0$ and $ J_0$ according to\cite{artigoRTN}
\begin{equation}
J^{\text{RTN}}_k(t,\tau_c)=(-1)^{f(t,\tau_c,k)} J_0,
\end{equation}
where $k=1,2$ and the function $f(t,\tau_c,k)$ is related to the times where the jumps occur by the following expression
\begin{equation}
f(t,\tau_c,k)=\sum_j\Theta(t-t^k_j),
\end{equation}
where $\Theta(x)$ is the Heaviside function and
\begin{equation}
t^k_j=-\sum_{n=1}^j\tau_c \;\log{(p^k_n)}.\label{tjk}
\end{equation}
In Eq.~(\ref{tjk}), $p^k_n$ are uniformly distributed random numbers and the correlation time $\tau_c$ determines the frequency of jumps and is related to the autocorrelation function as follows
\begin{equation}
\langle J^{\text{RTN}}_k(t,\tau_c)J^{\text{RTN}}_k(t',\tau_c)\rangle=\exp{(-2|t-t'|/\tau_c)}, \label{autocorr}
\end{equation}
where $\langle \cdots\rangle$ represents an average over the fluctuations.

\section{Results}
The results for the $J^{\text{RTN}}_1(t,\tau_c)$ are shown in Fig.~1, considering $J_0=1$ MHz and $\tau_c=10$ ns (black-solid curve in the top panel) and $\tau_c=30$ ns (red-dashed curve in the bottom panel). As expected, there are less jumps for a higher value of $\tau_c$. Furthermore, to check the 1/f$^\alpha$ nature of the RTN, we numerically calculate the power spectrum $S(f)$, which is the Fourier transform of the autocorrelation function (Eq.~(\ref{autocorr})). Such results are shown in Fig.~2 for a fixed amplitude $J_0=1$ MHz and for different correlation times: $\tau_c=1$ ns (magenta dotted curve), $\tau_c=10$ ns (black solid curve), and $\tau_c=30$ ns (blue dashed curve). The solid red curve in Fig.~2 is a plot of the function $4\times 10^{-3}/f^{0.89}$, which is used for comparison to the high frequency behavior of the numerically calculated power spectrum $S(f)$. Such a power-law noise model is similar to the one deduced for only one qubit in DQDs.~\cite{prl_shulman} 
\begin{figure}[t]
\includegraphics[width=8cm]{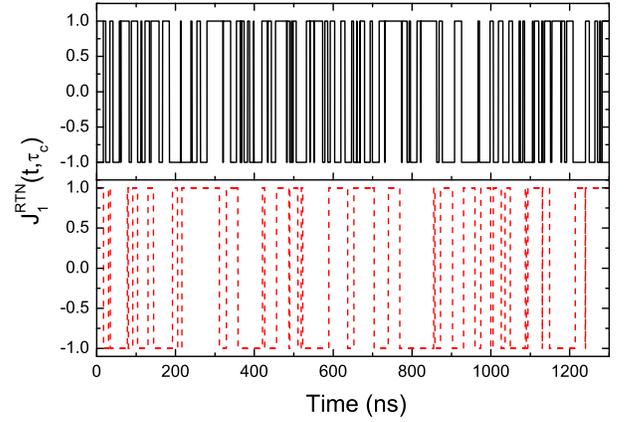}
\caption{\label{fig1} (Color online) Simulation of the RTN as a function of time, considering the amplitude $J_0=1$ MHz and two different correlation times: $\tau_c=10$ ns (top pannel) and $\tau_c=30$ ns (bottom pannel).}
\end{figure}

\begin{figure}[b]
\includegraphics[width=8cm]{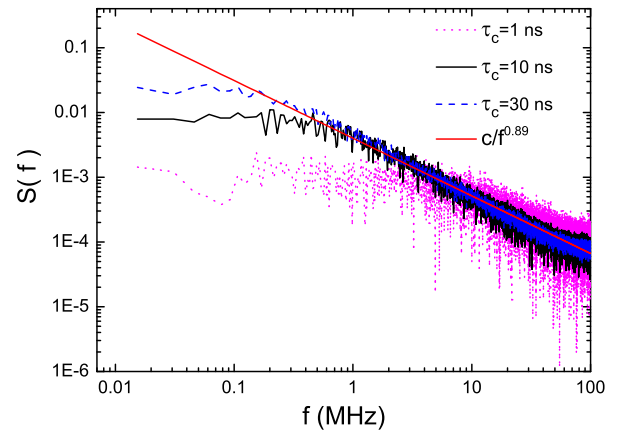}
\caption{(Color online) Numerical calculation of the power spectrum S(f) considering the RTN amplitude $J_0$ = 1 MHz and the following values for the correlation time: $\tau_c$ = 1 ns (magenta dotted curve),  $\tau_c$ = 10 ns (black solid curve), and $\tau_c$ = 30 ns (blue dashed curve). The function c/$f^0.89$ is plotted as a solid red curve for comparison.}\label{fig2}
\end{figure}

To perform the analysis of our results and to compare to the experimental work,~\cite{sci} the quantum correlation called concurrence is employed. Concurrence is a well known measure of entanglement, which is broadly accepted to be responsible for a set of important tasks in quantum information theory, such as quantum teleportation \cite{tele} and quantum key distribution.~\cite{key} For two qubits, there is an analytical solution to concurrence,~\cite{concurrence} which is given by
\begin{equation}
C(\rho)=\max\{0,{\lambda_1}-{\lambda_2}-{\lambda_3}-{\lambda_4}\},
\end{equation}
where $\lambda_i$ ($i=1,2,3,4$) are the eigenvalues of $R=\sqrt{\sqrt{\rho}\tilde\rho\sqrt{\rho}}$ listed in descending order. $\tilde{\rho}$
is the time-reversed density operator, which can be written as
\begin{equation}
\tilde{\rho}=(\sigma^{(1)}_y\otimes\sigma^{(2)}_y)\rho^*(\sigma^{(1)}_y\otimes\sigma^{(2)}_y),
\end{equation}
where $\rho^*$ is the conjugate of $\rho$ in the standard basis of two qubits.

The initial state of each DQD is set to $\mid\uparrow\rangle=\mid S\rangle$, then a $\pi/2$ rotation around the $x$ axis is performed during the preparation time $\tau_{prep}$, which puts each qubit in a superposed state $(\mid\uparrow\rangle+\mid\downarrow\rangle)/\sqrt{2}$. Following the experimental description given in ref.~[\onlinecite{sci}], we use $\Delta B_{z,1}=\Delta B_{z,2}= \pi/(2\tau_{prep})$, $J_1/2\pi = 280 {\rm {MHz}}$, and $J_2/2\pi = 320 {\rm {MHz}}$. The system dynamics can be obtained by numerically solving the unitary trajectories described by the system Hamiltonian (Eq.~(\ref{ham})) together with the RTN Hamiltonian (Eq.~(\ref{hamRTN})). We perform an average over different unitary trajectories to extract the dynamics of the system including the RTN.~\cite{numerical}

\begin{figure}[t]
\includegraphics[width=8cm]{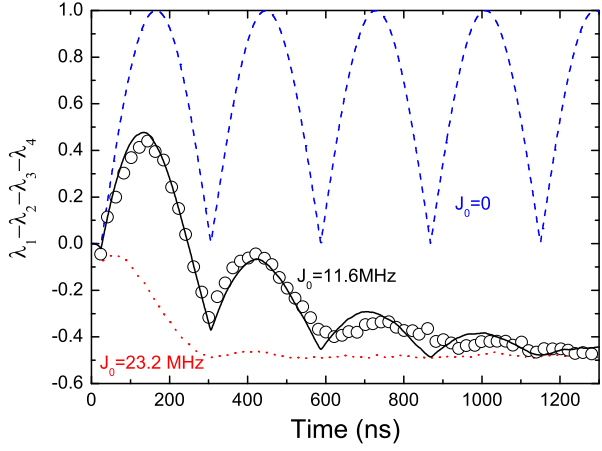}
\caption{\label{fig3} (Color online) Time-evolution of the DDSE considering the RTN with fixed correlation time  $\tau_c$ =9 ns and for different noise amplitudes: $J_0$ = 0 (dashed blue curve), $J_0$ = 11.6 MHz (solid black curve), and $J_0$ = 23.2 MHz (dotted red curve). Open circles denote the DDSEs extracted from experimental data.~\cite{sci}}
\end{figure}

\begin{figure}[t]
\includegraphics[width=8cm]{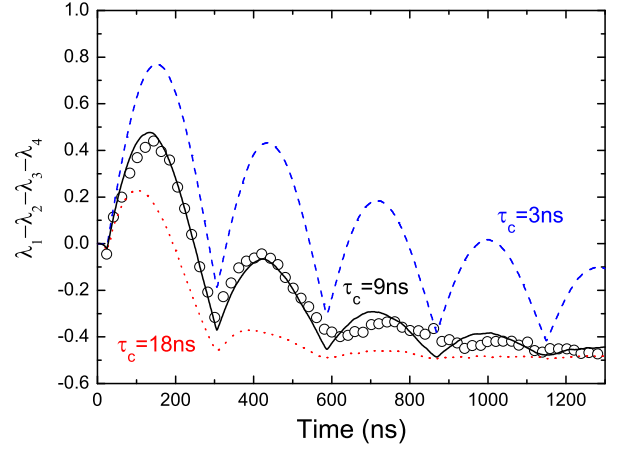}
\caption{\label{fig4}(Color online)Time-evolution of the DDSE considering the RTN with fixed noise amplitude $J_0$ = 11.6 MHz and for different values of the correlation time: $\tau_c$ =3 ns (dashed blue curve), $\tau_c$ = 9 ns (solid black curve), and $\tau_c$ = 18 ns (dotted red curve). Open circles denote the DDSEs extracted from experimental data.~\cite{sci}}
\end{figure}

We begin our analysis of the dynamics of the system through the evolution in time of the difference of the descending sorted eigenvalues (DDSE) ${\lambda_1}-{\lambda_2}-{\lambda_3}-{\lambda_4}$ of the matrix $R=\sqrt{\sqrt{\rho}\tilde\rho\sqrt{\rho}}$, which is equal to the concurrence $C(\rho)$ when it assumes positive values. 
In Fig.~(\ref{fig3}), we plot ${\lambda_1}-{\lambda_2}-{\lambda_3}-{\lambda_4}$ as function of time, assuming different values for the RTN amplitude $J_0$ and for $\tau_c=9$~ns. When $J_0=0$, the dynamics is unitary and the concurrence oscillates without dissipation with a period of $280$ ns after $\tau_{prep}$. Such a period is completely defined by the system Hamiltonian term ${J_{12}/4}\sigma_z^{(1)}\otimes\sigma_z^{(2)}$. DDSE assume zero values for $t\le \tau_{prep}$ when $J_0=0$, but the first experimental value of DDSE at $t=25$~ns is negative and it must be related to noise effects that occur during the preparation of the state $(\mid\uparrow\rangle+\mid\downarrow\rangle)/\sqrt{2}$. Such a negative value occurs during the preparation time as a result of the interaction of the $\pi/2$ rotation around the $x$ direction and the RTN, which acts in the $z$ direction. Indeed, to take into account such a negative value of DDSE, we just need to use $\tau_{prep}=25$~ns and $J_0\neq 0$, as can be observed in Fig.~(\ref{fig3}). Furthermore, one can see in Fig.~(\ref{fig3}) that the DDSE oscillates in time and it has an envelope function that decays faster as the value of $J_0$ is increased. We also perform an analysis of DDSE as a function of time for a fixed value of the RTN amplitude $J_0=11.6$ MHz and different correlation times $\tau_c$, which is shown in Fig.~(\ref{fig4})). The increasing of the correlation time has a similar effect when compared to the increasing of the RTN amplitude in Fig.~(\ref{fig3}); \textit{i.e.}, the bigger the correlation time, the faster the decay of the envelope function as a function of time. For $\tau_c=9$ ns and $J_0=11.6 \rm{MHz}$, there is a good matching between experimental results (open circles in Figs.~(\ref{fig3}) and (\ref{fig4}) and DDSE extracted from the dynamics including the RTN. Results shown in Figs.~(\ref{fig3}) and (\ref{fig4}) might suggest that it is possible to find different pairs of $\tau_c$ and $J_0$ that adjust the experimental data, but this conception is misleading because other pairs of $\tau_c$ and $J_0$ cannot reproduce the experimental data in whole range of time.

By means of our description of the decoherence mechanism, we can analyze the entanglement dissipative dynamics through our theoretical model. Particularly, we focus on two aspects: the role of the preparation time $\tau_{prep}$ and the two-qubit coupling $J_{12}$. The system is interacting with the environment during the preparation time, which affects the entanglement efficiency. To understand the role of such a physical parameter, we analyse  effects on the maximum value of concurrence caused by distinct preparation times $\tau_{prep}$. Another crucial physical parameter that rules the entanglement is the coupling between each qubit. The two-qubit coupling $J_{12}$, that can be increased by controlling the dipole-dipole interaction, determines the time $\tau_{ent}$ for achieving the state with highest value of entanglement (concurrence). This time $\tau_{ent}$ can be extracted from the term $J_{12}/4\sigma_z^{(1)}\otimes\sigma_z^{(2)}$ of Eq.~(1) and is given by $J_{12} = \pi/\tau_{ent}$. In the experimental results, $\tau^{exp}_{ent}=140~{\rm{ns}}$ and the maximum obtained value for entanglement is around $0.44$.~\cite{sci} In other words, both $\tau_{prep}$ and $\tau_{ent}$ are two fundamental characteristic times that are intrinsically related to the success of achieving an maximally entangled state.

\begin{figure}[t]
\includegraphics[width=8cm]{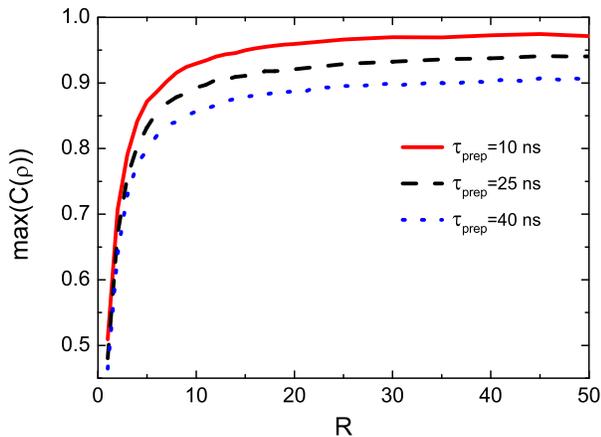}
\caption{\label{fig5} Numerical solution for the maximum value of the concurrence as a function of R and for $\tau_{prep}=10$~ns, $\tau_{prep}=25$~ns, and $\tau_{prep}=40$~ns, considering the description of the noise that better fits the experimental data, \textit{i.e.} $\tau_c=9$ ns and $J_0=11.6 \rm{MHz}$.}
\end{figure}

\begin{figure}[b] 
\includegraphics[width=8cm]{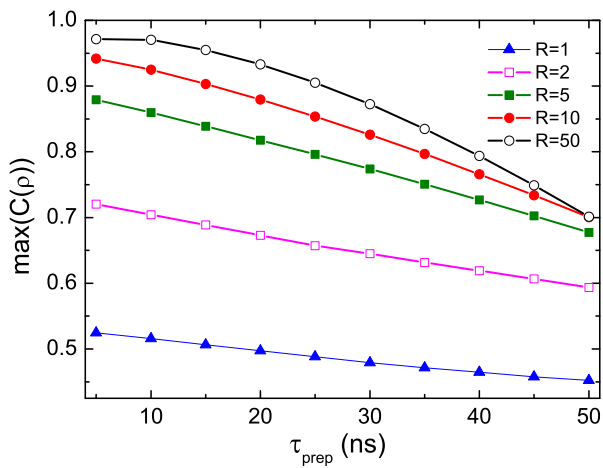}
\caption{\label{fig6} Numerical solution for the maximum value of the concurrence as a function of $\tau_{prep}$ for R=1, R=2, R=5, R=10, and R=50, considering the description of the noise that better fits the experimental data, \textit{i.e.} $\tau_c=9$ ns and $J_0=11.6 \rm{MHz}$.}
\end{figure}

To illustrate the role of the preparation time and the dependency on the two-qubit coupling $J_{12}$ in the entanglement dynamics, we plot in Fig.~(\ref{fig5}) the maximum value of concurrence as a function of $R=J_{12}/J^{exp}_{12}$, where $J^{exp}_{12}$ is the value extracted from the experimental data, for different preparation times $\tau_{prep}$. As expected, such results show an enhancement of the entanglement when $J_{12}$ is increased and when $\tau_{prep}$ is decreased. This behavior is related to the fact that a maximally entangled configuration is faster achieved for a larger $J_{12}$ even though the RTN disturbs the ideal obtainment of the superposed state during the preparation time $\tau_{prep}$. In Fig.~(\ref{fig5}), one can see that for $R\gtrsim 20$ the maximum value of entanglement is approximately constant.

Furthermore, a small increase in the value of $J_{12}$ surprisingly enhances the maximum value of entanglement, as can be observed by the steep jump in $\rm{max}(C(\rho))$ for a small variation of R in Fig.~(\ref{fig5}). To understand further how these characteristic times affect the maximally entangled state, in Fig.~(\ref{fig6}), we plot the maximum value of concurrence as a function of $\tau_{prep}$ for $R=1$, $R=2$, $R=5$, $R=10$, and $R=50$. One can notice in Fig.~(\ref{fig6}) that the maximum concurrence monotonically decreases as a function of $\tau_{prep}$ and that the maximum concurrence rapidly increases with the increasing of $R$. 

For example, by doubling the experimental value of the two-qubit coupling ($R=2$), the maximum value of entanglement has a growth of $40\%$ for $5\leq \tau_{prep}\leq 50$. These results address the way such physical parameters can be tuned in order to substantially enhance the entanglement between $S$-$T_0$ qubits.

\section{Conclusion}
In summary, we proposed a model based on the RTN, which mimics the 1/f$^\alpha$ noise, to describe the dissipative dynamics of two $S$-$T_0$ qubits in two DQDs. By employing such a model, we were able to determine a suitable description of the experimental data shown in ref.~[\onlinecite{sci}].  Moreover, we studied the role of the preparation time and the two-qubit coupling in the dissipative dynamics. We showed that the two-qubit coupling plays a crucial role in the entanglement evolution and a small increase in $J_{12}$ can lead to a considerable amplification of the entanglement. Such results can be used as a reference for further studies in quantum systems where the 1/f$^\alpha$ noise is present.

\section{Acknowledgements}
We thank M. Shulman and F. Brito for helpful discussions.
LKC and FFF are grateful to the Brazilian Agencies FAPESP (grants 12/13052-6 and 15/05581-7), CNPq (grants 	304841/2015-3 and 474592/2013-8), and CAPES for financial support. KB and FFF would like to thank the International Centre for Theoretical Physics (ICTP) for financial support.


\begin{thebibliography}{99}
\bibitem{numerical} In our numerical simulations, we used 3000 trajectories and a time step of $\Delta t=1$~ns for all results, except for the results of Fig. 2, where we used 5000 trajectories and a time step of $\Delta t=0.1$~ns.
\end{thebibliography}
\end{document}